\documentclass[preprint,superscriptaddress,showpacs,preprintnumbers,amsmath,amssymb,aps,longbibliography]{revtex4-1}

\usepackage[utf8]{inputenc} 
\usepackage{epsfig}
\usepackage{amsmath}
\usepackage{subfigure}
\usepackage{epstopdf} 
\usepackage{graphicx}
\usepackage{color}
\usepackage{pict2e}
\usepackage{mathtools}

\DeclarePairedDelimiter\abs{\lvert}{\rvert}

\newcommand{\comment}[1]{}

\begin{document}

\title{Average conservative chaos in quantum dusty plasmas}

\author{\'{A}lvaro G. L\'{o}pez}
\affiliation{Nonlinear Dynamics, Chaos and Complex Systems Group, Departamento de	
F\'{i}sica, Universidad Rey Juan Carlos, Tulip\'{a}n s/n, 28933 M\'{o}stoles, Madrid, Spain}

\author{Rustam Ali}
\affiliation{Department of Mathematics,Visva Bharati University, Santiniketan 731235, India}
\affiliation{Department of Mathematics, Sikkim Manipal Institute of Technology,Sikkim Manipal University, Majitar, Rangpo, East-Sikkim 737136, India}

\author{Laxmikanta Mandi}
\affiliation{Department of Mathematics,Visva Bharati University, Santiniketan 731235, India}
\affiliation{Department of Mathematics,Gushkara Mahavidyalaya,Purba Bardhaman 713128,India}

\author{Prasanta Chatterjee}
\affiliation{Department of Mathematics,Visva Bharati University, Santiniketan 731235, India}

\date{\today}

\begin{abstract}
We consider a hydrodynamic model of a quantum dusty plasma. We prove mathematically that the resulting dust ion acoustic plasma waves present the property of being conservative on average. Furthermore, we test this property numerically, confirming its validity. Using standard techniques from the study of dynamical systems, as for example the Lyapunov characteristic exponents, we investigate the chaotic dynamics of the plasma and show numerically its existence for a wide range of parameter values. Finally, we illustrate how chaotic dynamics organizes in the parameter space for fixed values of the initial conditions, as the Mach number and the quantum diffraction parameter are continuously varied.
\end{abstract}

\maketitle

\textbf{We demonstrate the chaotic feature of dust ion acoustic waves in a quantum dusty plasma model. We derive a four dimensional dynamical system of ordinary differential equations using the basic fluid equations and show that the dynamical system is conservative on the average. To detect chaotic waves in this system the maximum
Lyapunov exponent (MLE) is computed and plotted in
two dimensional parameter (quantum diffraction parameter, Mach number) space to identify regions where weak and not so weak chaos appears.}

\section{Introduction}

Quantum dusty plasmas are receiving increasing attention in the past recent years because of their numerous applications in the field of micro- and nano manufacturing \cite{craighead2000}. Consequently, the nonlinear structure (viz., solitary waves, shock structure, double layers and chaos) associated with waves in these plasmas have been the subject of extensive analysis. A plasma is a many-body system composed of a very large number of charged particles whose dynamics are dominated by long-range collective effects mediated by the electromagnetic force. Intuitively, it can be regarded as a quiasineutral electrically conductive fluid. In particular, dusty plasmas are low temperature plasmas consisting of electrons, ions, neutral particles and very massive micrometer-sized solid charged dust grains \cite{Rao90,barkan1996,Verheest00,wang2001,shukla2002i}. The grains are charged because they collect flowing electrons and ions. Since the electrons are considerably lighter, their higher average speeds lead to an imbalance in the collection of charges, rendering dust particles their negative charge. Then, the polarization of the surrounding space by these dust grains can lead to the shielding of charge, which limits the range of interaction of the dust particles (see Fig. 1).
\begin{figure}[ht!]
\centering
\includegraphics[width=0.8\textwidth]{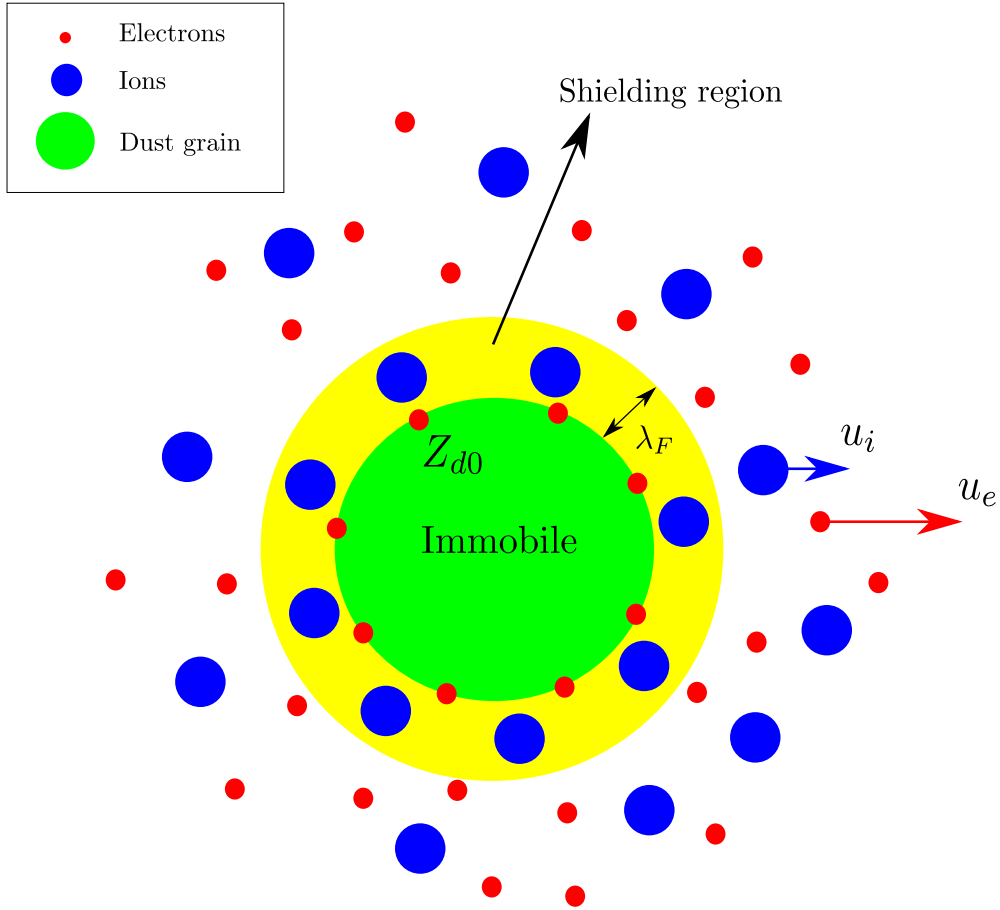}
 \caption{\textbf{Dusty plasma}. A system of charged electrons, ions and dust grains. These grains are very heavy in comparison to the ions and the electrons and, consequently, considered at rest in some inertial frame. In this frame, the ions and the electrons are flowing. Generally, ions tend to travel slower than electrons, due to their larger inertia. The flowing electrons and the ions tend to unequally attach to the grains, leading to a net negative charge on the dust surface. Then, these charges polarize the surrounding medium leading to a shielding region (yellow). In quantum dusty plasmas this region has a typical length called the Thomas-Fermi length $\lambda_F$.}
\label{fig1}
\end{figure}

In classical plasmas the de Broglie wave length 
\begin{equation}
\lambda_{B_\alpha}=\frac{\hbar}{\sqrt{m_\alpha k_B T}}
\end{equation}
plays no relevant role because of its smallness compared to the average interparticle distance $\bar{r}_{\alpha}= \propto n_\alpha^{-1/3}$, where $n_\alpha$ and $m_\alpha$ are the number density and mass of the $\alpha$-species, $k_B$ is the Boltzmann constant,  $\hbar$ is Planck's constant $(h)$ divided by $2\pi$ and $T$ is the system's temperature. However, in quantum plasmas, the quantum effect is taken into consideration mostly when the thermal de Broglie wave length is similar or larger than the average interparticle distance ($i. e.$ when $n_\alpha \lambda_{B_\alpha}\geq 1$). As the de Broglie wavelength depends upon the mass of the $\alpha$-species and on the thermal energy $k_B T$, and since the electron mass is much less than the mass of the ions, the quantum effects associated with the electrons are generally more important than that of ions. Equivalently, the quantum effect can not be neglected if the system's temperature is comparable to or lower than Fermi's temperature $T_F=E_F/k_B$, where
\begin{equation}
E_{F \alpha}=\frac{\hbar^2}{2m_{\alpha}}(3\pi^2 n_{\alpha})^{2/3}
\end{equation}
is Fermi energy of the species $\alpha$. Recall, the Fermi energy $E_{F}$ at zero temperature represents the energy of the last occupied level in a charged gas. Hence, the quantum effects become relevant as the ratio $T_F/T$ increases over one.

Waves of interconnected charged particles propagating in a more or less periodic way frequently appear in plasma physics. 
A signature of these waves is that their phase velocity is larger than the thermal velocities of the dust and the ions. The restoring force leading to the plasma oscillations comes from the electron pressure, while the inertia is mostly provided by the dust and the ions. These waves have been found to occur in many different modes, for example, the dust acoustic (DA) mode \cite{Rao90,verheest2005}, the dust drift mode \cite{shukla1991}, dust lattice and cyclotron modes \cite{melandso1996,merlino1998}, dust ion acoustic (DIA) mode \cite{shukla1992,nakamura1999,mamun2002,kourakis2004,mamun2009,mamun2011,mamun2015}, dust Berstain-Green-Kruskal mode \cite{Tribeche04}, and so on. Waves in dusty plasmas have been widely studied both theoretically and experimentally during the last few decades because these low frequency electrostatic waves have a wide range of applications in laboratory, space and laser plasma physics \cite{brau92,Weixing1993}. However, recent works suggest that dust particles cannot survive in quantum plasmas because the electrons produce high pressures that can destroy any micrometer and even nanometer sized particles. Therefore, and as can be inferred from the parameter values of the plasma provided in the model description, the conclusions of the present study should be restricted to laboratory produced quantum plasmas, at best \cite{molda2019}.

The first theoretical prediction of the existence of low frequency DIA waves in dusty plasma consisting of negatively charged static dust grains were made by Shukla \emph{et al.} \cite{Shukla92}. It did not take very long until these waves were experimentally observed in homogeneous unmagnetized dusty plasmas \cite{Barkan96,nakamura1999}. They observed that the phase velocity of the wave increases and the wave endures heavy damping with increasing dust density in the linear regime. Further progress on quantum plasmas investigated several basic features of cylindrical and spherical DIA solitary waves containing inertial ions, Boltzmann electrons and stationary dust particles in unmagnetized dusty plasma \cite{mamun2002}. Recent works have observed shock and solitary waves in dusty plasmas in the presence of Boltzmann electrons, Boltzmann negative ions, warm positive ions and charged stationary dust \cite{mamun2010}. These outstanding works have shown that dust charge fluctuations is a source of dissipation, which is responsible for the formation of DIA shock waves. Finally, the latest works have reported the quasiperiodic and chaotic nature of these waves as the Mach number and the quantum parameter $H$ are varied \cite{sahu2015}. Ghosh et al. \cite{ghosh2014} studied the existence of chaotic, quasiperiodic, and periodic structures of DIA waves for quantum dusty plasmas using a dynamical systems approach, while Banerjee et al. \cite{banerjee2018} studied DIA solitary waves using a quantum hydrodynamic model by means of a Sagdeev’s nonperturbative method.

The aim of the present work is to show the existence of average conservative chaos in nonlinear DIA waves using a one-dimensional quantum hydrodynamic model of an unmagnetized, collisionless, ultracold dusty plasma. For this purpose we consider the quantum mechanical behavior of electrons and neglect the quantum diffraction effects of ions because of their larger inertia \cite{hass2003,sahu2007,mushtaq2007}. Nevertheless, the statistical effects due to ions are taken into consideration to some extent by assuming some degeneracy ion pressure \cite{haas2005,stenflo2006,marklund2007}. Finally, we disregard exchange-correlation effects as well, which can have important consequences in a region where the electron density is high and the temperature is very low \cite{mebrouk2014,mahmood2019,ahmad2020}. In Sec. II we describe the model and the approximations in detail. Then, in Sec. III, analytical results demonstrating the average conservative nature of the plasma waves appearing in the model are provided. In Sec. IV we introduce the numerical methods used to detect the existence of chaotic dynamics, while in Secs. V and VI we carry out a systematic numerical study to gain insight into the chaotic nature of the resulting nonlinear plasma waves. Finally, we provide discussion concerning our results in relation to previous works and on the modeling of nonlinear plasmas.

\section{Model description}
A system consisting of electrons, positively charged ions, equiradius spherical dust grains carrying identical charge and mass is considered. Among the different  approaches to represent unmagnetized collisionless quantum plasmas, the quantum Dawson model uses an hydrodynamic description based on Madelung's picture of quantum mechanics \cite{haas2011quantum}. The governing equations used to study low phase velocity quantum DIA oscillations in a one-dimensional dusty plasma can be written as
\begin{eqnarray}
\frac{\partial n_i}{\partial t}+\frac{\partial(n_iu_i)}{\partial x}=0,\label{eq:3}~~~~~~~~~~~~~~~~~~~~~~~~~~~~~~~~~~~~~~~~~~~~\\ 
	\frac{\partial u_i}{\partial t}+u_i\frac{\partial u_i}{\partial x}=-\frac{e}{m_i}\frac{\partial \phi}{\partial x}-\frac{1}{n_im_i}\frac{\partial p_i}{\partial x}-\frac{1}{m_i}\frac{\partial Q_i}{\partial x},\label{eq:4}~~~~~~~~\\
    \frac{\partial u_e}{\partial t}+u_e\frac{\partial u_e}{\partial x}=\frac{e}{m_e}\frac{\partial \phi}{\partial x}-\frac{1}{n_em_e}\frac{\partial p_e}{\partial x}-\frac{1}{m_e}\frac{\partial Q_e}{\partial x},\label{eq:5}~~~~~~~~~\\
	\frac{\partial ^2 \phi}{\partial x^2}=4\pi e(n_e+Z_{d0}n_{d0}-n_i).\label{eq:6}~~~~~~~~~~~~~~~~~~~
\end{eqnarray}
where $n_i (n_e)$ is the number density of ion (electron), $u_i(u_e)$ is the ion (electron) fluid speed, $m_i (m_e)$ is the mass of the ion (electron) and $\phi$ is the electrostatic potential and electron (ion) charge is given by $-e(e)$. We briefly explain these equations. Eq.~\eqref{eq:3} represents the continuity equation for ions, which guarantees the conservation of the total number of charged particles $n_i$. A similar equation holds for the other species $n_e$ of this two stream model, but we shall not use it in the present study. Then, two eulerian fields governed by Eqs.~\eqref{eq:4} and \eqref{eq:5} describe the rate of change in the momentum $u_\alpha$ of ions and electrons, respectively. The internal forces appearing in these two equations are given, on the one hand, by the Fermi pressure of the plasma and, on the other, by the quantum potential of each species, which can be written as
\begin{equation}
Q_{\alpha}=-\dfrac{\hbar^2}{2 m_{a}}\frac{1}{\sqrt n_\alpha}\frac{\partial^2 \sqrt n_\alpha}{\partial x^2}.
\label{eq:7}
\end{equation}
This force involves the tunneling of degenerate electrons through the Bohm potential and, as has been recently suggested, also has an electromagnetic origin \cite{lopez2020}. The emergent average electromagnetic field of the plasma is represented by the gradient of the electrostatic potential $\phi$. If we disregard the degeneracy pressure, these momentum equations are tantamount to writing the Schr\"odinger equation for each species. Finally, the Poisson equation governs the average electromagnetic field of the plasma, which is assumed to be in instantaneous equilibrium. This quasistatic approach disregards the magnetic fields induced by Faraday's law, and is admissible as long as the fluctuations in the electric potential are not too fast.

Because of the heavier mass of the dust grains as compared to that of the electrons and the ions, their dynamics evolves on a much longer time scale, hence the dust grains are taken to be immobile and negatively charged $q_d=-Z_{d0} e$, where $Z_{d0}$ is the number of electrons residing on dust surface. Here we assume that electrons and ions follow the one-dimensional zero-temperature Fermi gas pressure law \cite{hass2003}, which reads
\begin{equation}
p_\alpha =\dfrac{1}{3}\dfrac{m_\alpha v_{F\alpha}^2}{n_{\alpha0}^2}n_{\alpha}^3\label{eq:8}.
\end{equation}
In this equation the magnitude
\begin{equation}
v_{F\alpha}=\sqrt{\dfrac{2k_BT_{F\alpha}}{m_\alpha}}
\label{eq:9}
\end{equation}
represents the electron and ion Fermi speed respectively, $k_{B}$ is the Boltzmann constant, and $T_{F \alpha}$ is the Fermi temperature of the species.  At equilibrium, the quasi-neutrality condition is given by $n_{i0}=n_{e0}+Z_{d0}n_{d0}$, where $n_{e0}$ and $n_{i0}$ are the equilibrium number densities of electrons and ions respectively, while $n_{d0}$ is the equilibrium number density of dust grains. Adopting the normalization
 $$\bar{n}_i=\frac{n_i}{n_{i0}}, \bar{n}_e = \frac{n_e}{n_{e0}}, \bar{u}_i = \frac{u_i}{c_s}, \bar{u}_e = \frac{u_e}{c_s}, \bar{\phi} = \frac{e\phi}{2k_BT_{Fe}}, \bar{t} = \omega_{pi} t, \bar{x} = x\frac{\omega_{pi}}{c_s}$$ where the constant
\begin{equation}
\omega_{pi}=\sqrt{\frac{4\pi n_{i0}e^2}{m_i}}
\label{eq:10}
\end{equation} 
is the ion plasmon frequency and we have introduced the quantum ion-acoustic velocity
\begin{equation}
c_s=\sqrt{\frac{2k_BT_{Fe}}{m_i}}.
\label{eq:11}
\end{equation} 
Renaming variables without dashes, the Eqs.~(\ref{eq:3})-(\ref{eq:6}) can be rewritten as
 \begin{eqnarray}
 \frac{\partial{n_i} }{\partial t}+\frac{\partial(n_iu_i)}{\partial x}&=&0,\label{eq:12}\\
  \frac{\partial u_i }{\partial t}+u_i\frac{\partial u_i}{\partial x}+\frac{\partial \phi}{\partial x}+\rho n_i \frac{\partial n_i }{\partial x }&=&\frac{m_e}{m_i}\frac{H^2}{2}\frac{\partial}{\partial   x }\bigg(\frac{1}{\sqrt{ n_i }}\frac{\partial^2\sqrt{ n_i }}{\partial x^2}\bigg),\label{eq:13}\\
   \frac{m_e}{m_i}\bigg(\frac{\partial u_e }{\partial  t}+ u_e\frac{\partial u_e }{\partial x }\bigg)&=&\frac{\partial \phi }{\partial x}- n_e \frac{\partial  n_e }{\partial x}+\frac{H^2}{2}\frac{\partial}{\partial  x }\bigg(\frac{1}{\sqrt{ n_e}}\frac{\partial^2\sqrt{ n_e}}{\partial  x^2}\bigg)\label{eq:14}\\
   \frac{\partial ^2 \phi}{\partial x^2}&=&-n_i+N_d+\epsilon_i n_e\label{eq:15}.
 \end{eqnarray}
 
Here $\epsilon_{i}=n_{e0}/n_{i0}$ is the ratio between the unperturbed electron and ion densities. This parameter takes values between zero and one, and should be close to the former for ion-acoustic waves. The parameter $N_d=Z_{d0}n_{d0}/n_{i0}$  represents the ratio between the unperturbed dust density and the unperturbed ion density, and in the present study ought to be smaller than one and not far from it. Then, the ratio $\rho=T_{Fi}/T_{Fe}$ corresponds to the ratio between ion Fermi temperature and electron Fermi temperature. For ion acoustic waves, the speed of the electrons is greater than the phase velocity of the acoustic waves, while the ion speed is frequently smaller.  Consequently, $\rho$ is expected to be close to zero. Just to set some reference values, for plasmas in semiconductor quantum wells \cite{ahmad2020} typical parameter values can be taken as $n_{e 0} = 4 \times 10^{16} cm^{-3}$, $n_{i 0} = 5 \times 10^{16} cm^{-3}$, $n_{d 0} = 10^{13} cm^{-3}$, for the electron, ion and dust densities, respectively. These values lead to a ratio $Z_{d 0} = 10^{3}$, which allows to estimate the order of $N_d$ in the forthcoming section. Then, the Fermi temperature of electrons takes an approximate value of few degrees Kelvin $T_{F e} = 5 K$, while the temperature of each species can be estimated from this value as $T_{e} = 10~T_{F e}$, $T_{i} = 0.5~T_{e}$, and $T_{d} = 10^{-4}~T_{e}$. Finally, as it is well-known, the masses of ions and dust can be written as $m_{i} = 1.67 \times 10^{-24} g$, $m_{e}=9.1 \times 10^{-28} g$, and $m_{d} = 10~m_{i}$, respectively.

The nondimensionalized quantum parameter is defined as 
\begin{equation}
H=\sqrt{\dfrac{\hbar^2\omega_{pi}^2}{m_em_ic_{s}^4}}.
\label{eq:16}
\end{equation} 
This parameter is of key importance in our study, since it represents the ratio of the quantum energy due to the plasma oscillations in relation to the kinetic energy of the ions and the electrons. It governs the quantum effects of the system, which tend to dominate as $H$ increases above the value of one. As ions are two thousand times heavier than the electrons ($m_e/m_i \ll 1$), in a first approximation, we can neglect these terms. Therefore, disregarding the right hand side of Eq.~\eqref{eq:8} and the left hand side of Eq.~\eqref{eq:9}, the following reduced model is obtained 
\begin{eqnarray}
\frac{\partial n_i}{\partial t}+\frac{\partial(n_iu_i)}{\partial x}=0,\label{eq:17}
~~~~~~~~~~~~~~~~~~~~~~~~~~~~~~\\ 
	\frac{\partial u_i}{\partial t}+u_i\frac{\partial u_i}{\partial x}+\frac{\partial \phi}{\partial x}+\rho n_i\frac{\partial n_i}{\partial x}=0,\label{eq:18}~~~~~~~~~~~~\\
	\frac{H^2}{2}\frac{\partial}{\partial x}\left(\frac{1}{\sqrt n_e}\frac{\partial^2 \sqrt n_e}{\partial x^2}\right)-n_e\frac{\partial n_e}{\partial x}+\frac{\partial \phi}{\partial x}=0,\label{eq:19}\\
	\frac{\partial ^2 \phi}{\partial x^2}+n_i-N_d-\epsilon_i n_e=0.\label{eq:20}~~~~~~~~~~~~~~~~~~~~
\end{eqnarray}  

In order to study DIA waves of arbitrary amplitude, we introduce the transformation $\xi=x-ct$ to a comoving frame with Mach number $c$, which describes the velocity of the nonlinear wave structure. Upon substituting the above transformation in Eqs.~\eqref{eq:17}-\eqref{eq:20} and integrating these equations, imposing the conditions $n_i\rightarrow 1$,$\phi\rightarrow 0$, $u_i\rightarrow 0$ at some point $\xi$, usually far from zero, we obtain the following set of equations 
\begin{eqnarray}
n_i=\frac{c}{c-u_i},\label{eq:21}~~~~~~~~~~~~~~~~~~~~~~~~\\
\phi=\frac{\rho}{2}(1-n_i^2)+\frac{1}{2}(2cu_i-u_i^2),\label{eq:22}
\end{eqnarray}
together with the two differential equations governing the profile of the waves, which read
\begin{eqnarray}
\frac{d^2 \sqrt{n_{e}}}{d\xi^2}=\frac{\sqrt{n_e}}{H^2}(n_e^2-K-2\phi),\label{eq:23}~~~~~~\\
\frac{d^2\phi}{d\xi^2~}=\epsilon_in_e+N_d-n_i,\label{eq:24}~~~~~~~~~~~~
\end{eqnarray}
where the constant of integration $K$ has been introduced. This constant is related to the value of $n_e$ and its second derivative at the point $\xi$ where the previous conditions were given. Following other works  \cite{hass2003,ghosh2014}, we consider a value $K=1$, which is consistent with the parameter values of our plasma. Renaming variables as $n_e=A^2$ and eliminating $\phi$ from the Eqs.~\eqref{eq:21}-\eqref{eq:24}, we obtain the following set of ordinary differential equations
\begin{eqnarray}
\dfrac{d^2A}{d\xi^2}&=&\dfrac{A}{H^2}\left(A^4-1-\rho-2cu_i+u_i^2+\rho\left(\dfrac{c}{c-u_i}\right)^2\right),\label{eq:25}\\
\dfrac{d^2u_i}{d\xi^2}&=&\dfrac{(c-u_i)^3}{(c-u_i)^4-c^2\rho}\left(\left(1+\dfrac{3c^2\rho}{(c-u_i)^4}\right)\left(\dfrac{du_i}{d\xi}\right)^2+N_d+\epsilon_i A^2-\dfrac{c}{c-u_i}\right)\label{eq:26}.
\end{eqnarray}
Finally, after making the replacements $x_1=u_i$, $x_2=du_i/d\xi$, $x_3=A$ and $x_4=dA/d\xi$ into the Eqs.~\eqref{eq:25}-\eqref{eq:26}, we can rewrite these equations in the form of a four dimensional dynamical system, as
\begin{equation}
	 \begin{array}{llll}
	 \dfrac{d x_1}{d \xi}=x_2,\\
	 \dfrac{d x_2}{d \xi}=\dfrac{(c-x_1)^3}{(c-x_1)^4-c^2\rho}\left(x_2^2+\dfrac{3c^2\rho x_2^2}{(c-x_1)^4}+N_d+\epsilon_i x_3^2-\dfrac{c}{c-x_1}\right),\\
	 \dfrac{d x_3}{d \xi}=x_4, \\
	 \dfrac{d x_4}{d \xi}=\dfrac{x_3}{H^2}\left(x_3^4-1-\rho-2cx_1+x_1^2+\rho \left(\dfrac{c}{c-x_1}\right)^2\right).
	 \end{array}
	 \label{eq:27}
	 \end{equation}
	 
We notice that the symmetry $(x_3,x_4) \rightarrow (-x_3,-x_4)$ is present in this system. This symmetry is expected because the physical relevant variable is $x_3^2$, which represents the probability density. Therefore, any change in the sign of $\sqrt{n_e}$ leaves invariant the probability density and can be absorbed in the phase of the wave function.

\section{Average conservative systems}

In the present section we provide an analytical proof demonstrating that the system of differential equations governing the profile of the waves, which is represented by Eqs.~\eqref{eq:27}, is conservative on average. Given a flow $\Phi_\xi(x)$ defined on $\mathbb{R}^n$, we define this property by requiring that $\langle \nabla \cdot F \rangle=0$, where the average is performed over the one-parameter $\xi$ along its whole domain, which in our case are the reals $[0,\infty)$. The field $F(x)$ represents the infinitesimal generators of the flow, which for the cold quantum dusty plasma are given by the functions
\begin{equation}
	 \begin{array}{llll}
	 F_1(x)=x_2,\\
	 F_2(x)=\dfrac{(c-x_1)^3}{(c-x_1)^4-c^2\rho}\left(x_2^2+\dfrac{3c^2\rho x_2^2}{(c-x_1)^4}+N_d+\epsilon_i x_3^2-\dfrac{c}{c-x_1}\right),\\
	 F_3(x)=x_4, \\
	 F_4(x)=\dfrac{x_3}{H^2}\left(x_3^4-1-\rho-2cx_1+x_1^2+\rho \left(\dfrac{c}{c-x_1}\right)^2\right)\label{eq:28}.
	 \end{array}
\end{equation}
Let $D_{t}\subseteq\mathbb{R}^4$ represent the evolution of the flow $\Phi_\xi(x)$ of the system of Eqs.~\eqref{eq:28} from the initial domain $D_{t_0}\subseteq\mathbb{R}^4$ as time progresses from $t_0$ to $t$. Then, the system is said to be dissipative, conservative or expansive, according to the value $\nabla \cdot F <0$, $=0$ or $>0$, at a certain point in $D_t$. In particular, the divergence of our system at a specific point $x$ can be computed as  
\begin{equation}
\nabla\cdot F=\frac{\partial F_1}{\partial x_1}+\frac{\partial F_2}{\partial x_2}+\frac{\partial F_3}{\partial x_3}+\frac{\partial F_4}{\partial x_4}=\frac{2(c-x_1)^4x_2+6c^2\rho x_2}{(c-x_1)^5-c^2\rho(c-x_1)}.\label{eq:29}
\end{equation} 
Note that $\nabla \cdot F$ depends on the variables $x_1,~ x_2$ describing the ion's dynamics and the parameters $\rho, c$. This suggests the impossibility of discriminating the conservative or the dissipative nature of the plasma waves in general. Simply put, the ODE system can present net contraction at some time and its overall dynamics might be expansive at some other. The Eq.~\eqref{eq:29} can be written more simply as
\begin{eqnarray}
\nabla\cdot F&=&\dfrac{d}{dx_1} \log \bigg(\frac{(c-x_1)^3}{(c-x_1)^4-\rho c^2}\bigg)^2x_2, \label{eq:30}
\end{eqnarray} 
which yields the expression
\begin{eqnarray}
\nabla\cdot F&=&\frac{d}{d\xi}\log \left(\frac{(c-x_1)^3}{(c-x_1)^4-\rho c^2}\right)^2.\label{eq:31}
\end{eqnarray} 

From Eq.~\eqref{eq:29} one can obtain
\begin{eqnarray}
\langle \nabla\cdot F \rangle=\lim\limits_{X\rightarrow \infty}\frac{1}{X}\int_{0}^{X} \frac{d}{d\xi}\log\bigg(\frac{(c-x_1)^3}{(c-x_1)^4-\rho c^2}\bigg)^2 d\xi,\label{eq:32}
\end{eqnarray}
which immediately allows us to write
\begin{equation}
\langle \nabla\cdot F \rangle=\lim\limits_{X\rightarrow \infty}\frac{1}{X} \log\bigg(\bigg(\frac{c-x_{1X}}{c-x_{10}}\bigg)^3\frac{(c-x_{10})^4-\rho c^2}{(c-x_{1X})^4-\rho c^2}\bigg)^2.
\label{eq:33}
\end{equation}
If we further assume only finite wave profiles, the boundedness $x_1<c$ of the phase space trajectory implies that there must exist some finite $K$ such that
\begin{equation}
\abs*{\log\bigg(\bigg(\frac{c-x_{1X}}{c-x_{10}}\bigg)^3\frac{(c-x_{10})^4-\rho c^2}{(c-x_{1X})^4-\rho c^2}\bigg)^2} <K.
\label{eq:34}
\end{equation}
Thus, for a bounded orbit we have the result $\langle \nabla\cdot F \rangle=0$, which proves that the dynamical system represented in Eq.~\eqref{eq:28} is conservative on average whenever a trajectory is bounded, as previously claimed.
\begin{figure}[ht!]
\centering
\includegraphics[width=0.8\textwidth]{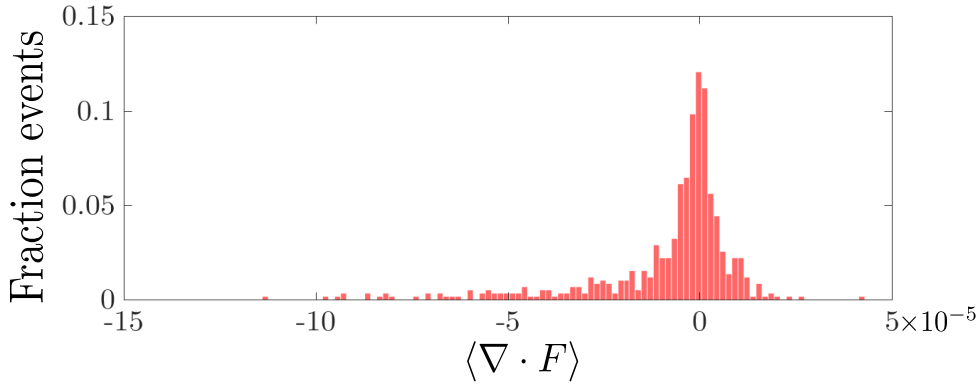}
 \caption{\textbf{Average conservative system}. A total number of $5 \times 10^4$ particles with randomly chose initial conditions are launched and evolved according to the Eqs.~\eqref{eq:28}. The average trace of the Jacobian is computed for each of them, and the fraction of cases is represented. As can be seen, most of the cases are around a value of zero, what confirms the average conservative nature of the waves, as defined above.}
\label{fig2}
\end{figure}

To numerically test this analytical result, we have considered solutions of the system of Eqs.~\eqref{eq:28}. Using parameter values described ahead, and by randomly selecting $5 \times 10^4$ initial conditions in the domain of interest, an histogram has been computed showing the frequency of different values of the average trace of the Jacobian. Certainly, the value is distributed around zero, with an approximately $70 \%$ of the events enclosed within a width of $1 \times 10^{-5}$. Some fat tail events appear, and some skewness can be detected as well, both presumably related to the complicated nature of the dynamical system at investigation and the simplicity of the integration scheme.

\section{Methods for chaos detection}

There exist a vast repertoire of numerical tools that allow to detect and characterize chaotic dynamics \cite{kath1997}. Certainly, the most used traditional chaotic indicator is the spectrum of Lyapunov characteristic exponents \cite{skokos2010}. As it is well-known, the whole spectrum allows to give bounds or compute other quantities of relevance, as for example the Kolmogorov-Sinai entropy \cite{benettin1976} or the Kaplan-Yorke dimension, which is useful when dissipative systems are being investigated and chaotic attractors appear. When interested only in a numerical proof of chaos, the maximal lyapunov exponent (MLE) is the most widely used method, for its simplicity and robustness.

Nevertheless, some fast algorithms have been developed recently, which do not require to compute the whole time series to obtain information from the system, which depend on the rate of convergence of the exponents. In this respect, the orthogonal fast Lyapunov indicators (OFLIs) are of great use \cite{barrio2005}, since they allow to unveil many features underlying the chaotic structure of the dynamical system, without relying on intensive high-time consuming computational resources. Finally, alignment indexes (SALI and GALI) deserve notification as well, which are mostly used in Hamiltonian systems. These methods are very efficient when studying the global dynamics of the system, and allow to discriminate in a simple way regular motion on low dimensional tori from the unpredictable trajectories of the surrounding chaotic sea \cite{Skokos2016}.

When studying the routes that lead to chaotic dynamics, bifurcation diagrams and the chaotic parameter set are two tools of fundamental importance \cite{kath1997}. These sets allow to visualize at a glance how chaos surges, evolves and organizes as some relevant parameters of the system are varied. In the present work, since we are dealing with a dynamical system that is only conservative on average, and since we mostly aim at showing the existence of chaotic waves as well as describing their organization in the parameter space, we shall use the more traditional Lyapunov spectrum and the MLE. This last indicator will be of great assistance to compute the \emph{chaotic parameter set}, which allows to see how chaotic regions embed in a frequently periodic or quasiperiodic parametric region \cite{gallas1993}.

The Lyapunov exponents are some of the oldest numerical detectors of chaos \cite{oseledec1968}. For a  orthonormal basis $\{v_i\}$ that spans the tangent plane of a point $x(\xi_0)$ in the phase space where the flow unfolds, we can define these exponents as
\begin{equation}
\lambda_i(x(\xi_0))=\lim_{\xi \to \infty} \frac{1}{\xi-\xi_{0}} \log ||M(\xi,\xi_0)v_i(\xi_0)||
\label{eq:35}
\end{equation}
where the matrix $M(\xi,\xi_0)=D\Phi_{\xi}(x(\xi_0))$ is the jacobian of the flow when the initial conditions are varied, assuming that the flow has evolved from $x(\xi_0)$ to $x(\xi)=\Phi_\xi(x(\xi_0))$. We note that the Lyapunov exponents generally depend on $x(\xi_0)$, except for ergodic systems. This matrix evolves according to the system of differential equations
\begin{equation}
\dfrac{d M(\xi,\xi_0)}{d \xi}=J(x(\xi)) M(\xi,\xi_0)
\label{eq:36}
\end{equation}
where $J(x)=D F(x)$ is the jacobian matrix of $F$, which defines the variational equations that govern how a small variation (a tangent vector) evolves under the flow until time $\xi$. Thus, these numbers are the result of computing the average rate of divergence or convergence of nearby trajectories in the phase space as time goes by. Simply put, the Lyapunov exponents measure the rates of growth or decrease of generic perturbations performed from a certain dynamical state as time goes by. Interestingly, in the present system we note that the average trace of the Jacobian gives $ \langle \nabla \cdot F \rangle = \lambda_1+\lambda_2+\lambda_3+\lambda_4$. In other words, the addition of Lyapunov exponents measures the rate of change of a volume element in phase space. Since our system is conservative on average, and one Lyapunov exponent is equal to zero, the one in the direction of the flow \cite{skokos2010}, we expect that two Lyapunov characteristic exponents are equal to zero, while the other two have opposite signs, because they must add to zero.

Among the  many algorithms available for computing Lyapunov exponents of a dynamical system, we shall use the most standard techniques \cite{benettin19801,Brown1991,Christiansen1997} that are known to calculate them. The whole Lyapunov spectra is computed by means of the traditional methods relying on the Gram-Schmidt orthogonalization procedure \cite{benettin19802}, while the MLE is computed also using the variational equations, but without normalizing the whole basis, but just one vector, since the largest exponent tends to dominate the others. To close this section, we recall that a fourth order Runge-Kutta integrator will be used all along the work. Other more accurate integrators with adaptive step size have been tested, showing very similar results. Thus we keep algorithms as simple as possible, since some computations are rather time-consuming, specially concerning the chaotic parameter set.

\section{Wave profiles and phase space}

Due to the property of being conservative on average, the system of Eqs.\eqref{eq:28} has an abstract interest on its own and deserves numerical exploration. Here we are mainly interested on understanding if chaotic dynamics occurs, and how it actually occurs in this dynamical system. Nevertheless, we give a detailed interpretation of the behaviour in terms of the plasma dynamics for the sake of completeness. 

In the present study we have used only the values of the plasma at one point as a condition to derive our equations. However, in other physical situations, where the whole boundary value problem is considered, only those solutions of the phase space that are compatible with the boundary conditions should be considered as physically acceptable. For example, if we have some fixed boundary condition representing equilibrium at $\xi \rightarrow \pm \infty$, an homoclinic trajectory beginning and ending its journey at the fixed point would have to be chosen as the solution to the ODE system, if it exists. In other words, when replacing a boundary value problem with a problem of initial conditions, as it is sometimes done when using dynamical systems theory as an approach to the study nonlinear waves \cite{hass2003,ghosh2014}, the results have to be considered with care, and sometimes should only be regarded as useful approximations that allow to investigate general features about the plasma dynamics, but never for practical applications.

We begin by computing the time series of the electron density $n_e$ and the ions speed $u_i$. Since $n_i=c/(c-u_i)$, the wave profiles of $n_i$ and $u_i$ are interrelated. These computation will allow us to unveil the structure and nature of some of these complex waves at a glance. In the second place, we compute Poincar\'e sections and their projections, which evince that chaos is rather weak for most of the series here displayed. As a first scenario we assume a quantum relative Mach number of $c=0.75$, thus the phase speed of the wave is $75\%$ of the quantum ion-acoustic velocity. The quantum diffraction parameter is set to $H=1.6$, what indicates that quantum effects are of importance. As previously suggested, we further take a small relation between the Fermi temperatures $\rho=0.09$, and take $N_d=0.98$ while recalling that $\epsilon_i=1-N_d$, which shows that the density of electrons at equilibrium represents $2\%$ of the charge contribution of the ions. This is typical of DIA waves, since the dust can collect a great deal of electrons \cite{shukla1992}.
\begin{figure}
\centering
\includegraphics[width=0.8\textwidth]{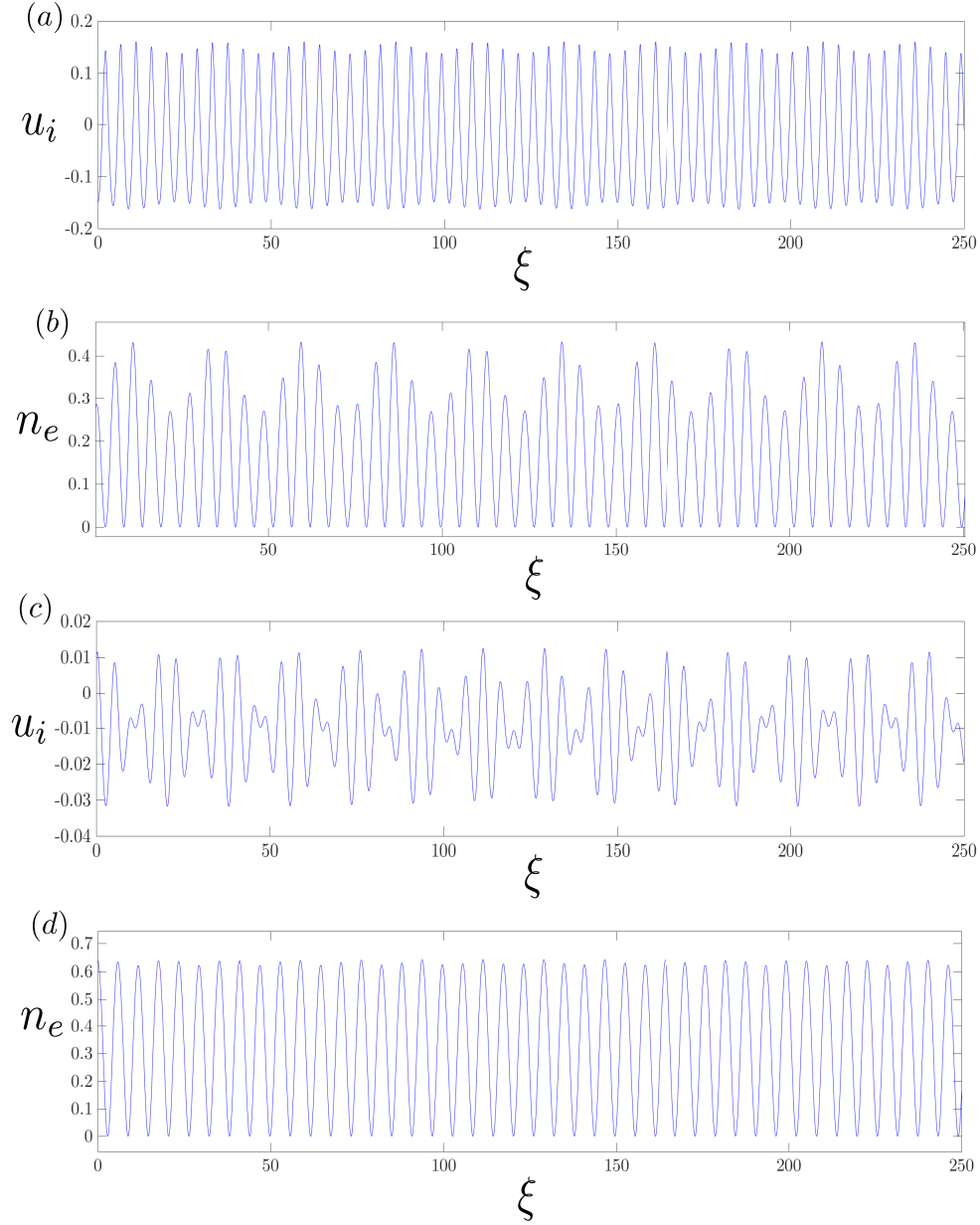}
 \caption{\textbf{Wave profiles}. Two waves for parameters $c=0.75$, $H=1.6$, $\rho=0.09$ and $N_d=0.98$ are shown. The first one considers $x_1=-0.15$, $x_2=0.00$, $x_3=0.50$ and $x_4=0.10$ as conditions at $\xi=0$, while the second wave assumes $x_{1}=0.01$, $x_{2}=0.01$, $x_{3}=0.80$ and $x_4=0.00$. (a) The ion speed for the first wave describes an apparently quasiperiodic motion, with small amplitude oscillations. (b) The electron density also showing profile with apparently periodic oscillations in amplitude, which follow the ion variations. (c) The second wave shows slow chaotic oscillations of the ion speed and density. (d) Very small fluctuations of the electron amplitude waves induced by the chaotic ion dynamics, showing an otherwise periodic wave.}
\label{fig3}
\end{figure}

As we can see in Figs.~\ref{fig3}(a) and (b), if at $\xi=0$ we set the conditions $x_1=-0.15$, $x_2=0.00$, $x_3=0.50$ and $x_4=0.10$, we have that the speed of the ions describe a wave with two discernible time scales. Both the phase and the amplitude seem to fluctuate with an apparently periodic fashion, with a group velocity that is approximately six times the value of the phase velocity. Nevertheless, imperfections appear in the profiles when we look at them more carefully. As we shall see bellow, these waves are neither periodic nor quasiperiodic according to our simulations, even though their Lyapunov exponents are very small. Noticeably, in those regions where the speed of the ions increases, electrons tend to accumulate as a consequence of the modification of the electrostatic potential. The fact that the electron density drops to zero is a consequence of neglecting their inertia in our approximation. However if we set the conditions $x_{1}=0.01$, $x_{2}=0.01$, $x_{3}=0.80$ and $x_4=0.00$ at $\xi=0$, we find a very different situation. Small perturbations from round values are sometimes allowed to test for robustness, but shall not be given special relevance by the reader. Here a background of ions are slowly swinging in an evidently chaotic fashion. These fluctuations in the ions speed and, therefore, in the ion density (see Eq.~\eqref{eq:21}), produce accordingly small chaotic fluctuations in the amplitude of the electron waves, for the same reasons as before.
\begin{figure}[ht!]
\centering
\includegraphics[width=0.9\textwidth]{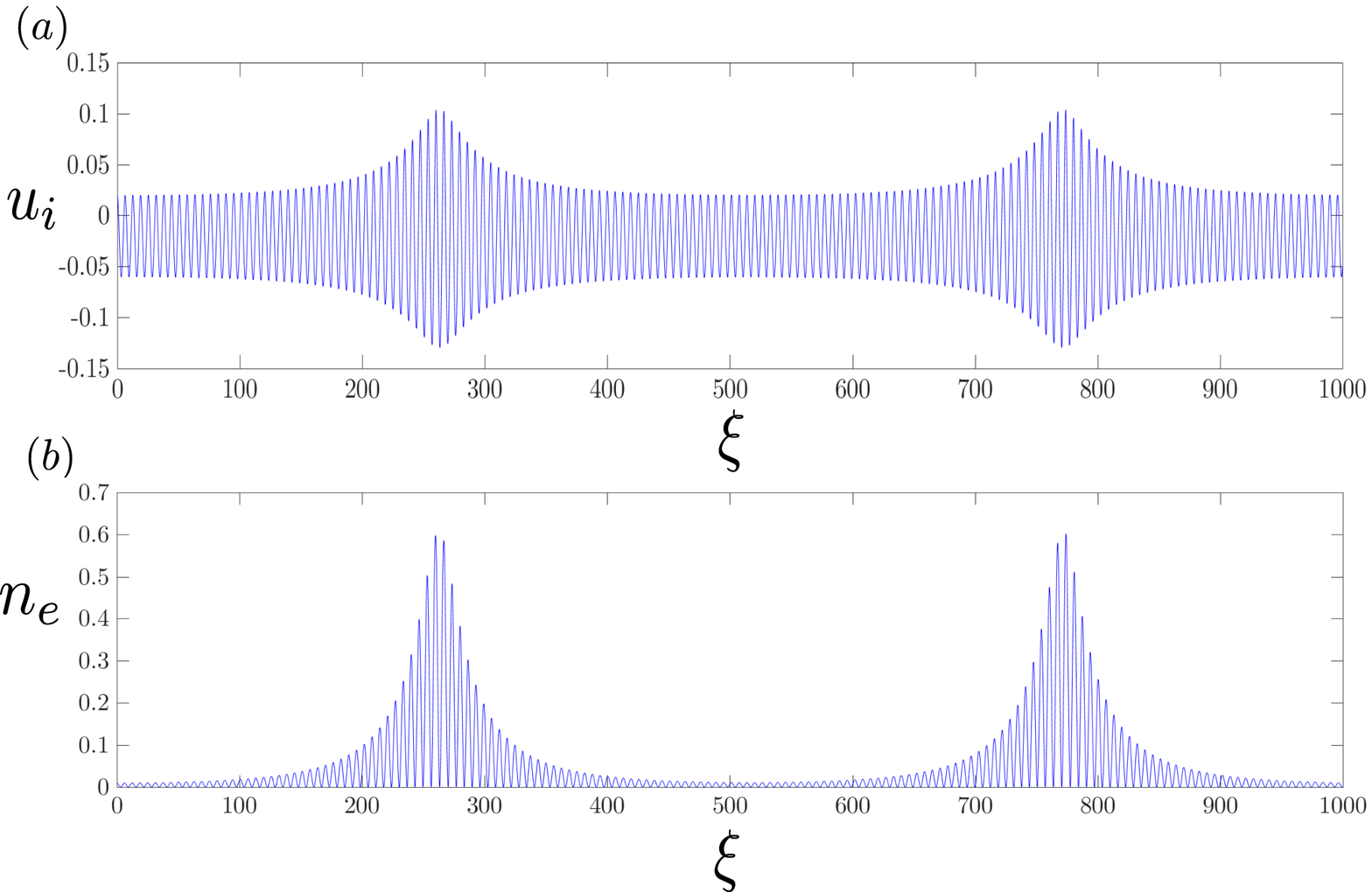}
\caption{\textbf{Wave profiles}. A plasma wave for parameters $c=1.0$, $H=2.0$, $\rho=0.05$ and $N_d=0.98$ is shown. We consider  $x_{1}=0.02$, $x_{2}=0.0$, $x_{3}=-0.10$ and $x_4=0.00$ as conditions at $\xi=0$. (a) The ion speed for the first wave describes a motion with fluctuations of the ion amplitude over very extended regions of space or time. (b) The electron density also showing a profile where the density of electrons is pronouncedly depleted from those regions where the amplitude of the ion density is constant.}
\label{fig4}
\end{figure}

As a second example we increase the quantum effects of the plasma to ascertain if the results previously obtained are reinforced by these new choice of parameters. These new values are achieved by considering a higher density of ions at equilibrium $n_{i0}$. In particular, we take a value $H=2.0$, which is sufficient to show the reinforcement of the effect. We also raise the Mach number to the value of the quantum ion acoustic velocity $c=1.0$ and we have reduced even further the relation between the Fermi speeds of the ions and the electrons $\rho=0.05$, just to make sure that the system is not very sensitive to variations in this parameter. Finally, the remaining parameter, which gives the ratio of dust grains to the electrons is kept equal, by setting $N_d=0.98$. The initial conditions at $\xi=0$ are set to $x_1=0.02$, $x_2=0.0$, $x_3=-0.1$ and $x_4=0.0$. As we can see in the Figs.~\ref{fig4}(a) and (b), this phenomenon is very similar to the one that appeared before, but much more pronounced. It is certainly interesting to see how the average acceleration of ions and their subsequent accumulation in certain regions of space attracts all the electrons, depleting them from widespread zones along the plasma. This phenomenon relies on the quantum potential and a possible explanation to it can be given as follows. Since the electron density in the plasma is already small in comparison to the ions, insofar as the dust grains adhere to their surface a great number of electrons, the concentration of these electrons in certain regions due to the electric field, leaves comparably very few electrons in the remaining zones of the plasma. Note that the width of the zones where the electrons accumulate are eight times smaller than the depleted zones, which leaves vast regions of the plasma occupied mostly by dust and ions.
\begin{figure}[ht!]
\centering
\includegraphics[width=0.8\textwidth]{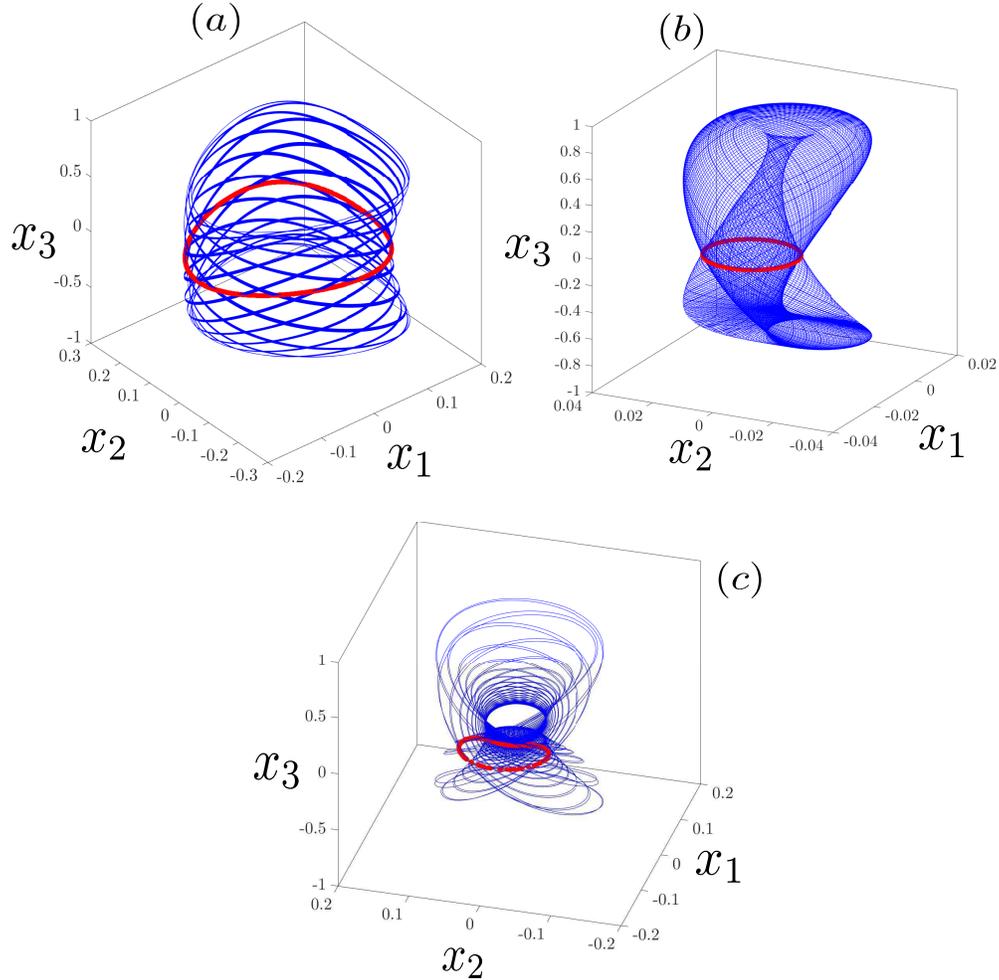}
 \caption{\textbf{Trajectories in phase space}. Three trajectories (blue) are projected onto the hyperplane $x_1x_2x_3$, together with points of the intersection of the trajectory, which have been projected as well. Some red points do not seem to belong to the blue trajectory, since these trajectories have not been fully represented, for clarity. (a) A trajectory with initial conditions $x_1=-0.15$, $x_2=0.00$, $x_3=0.50$ and $x_4=0.10$ at $\xi=0$, in the first parameter setting. (b) A trajectory with initial conditions $x_{1}=0.01$, $x_{2}=0.01$, $x_{3}=0.80$ and $x_4=0.00$ in the first parameter setting. (c) A trajectory with initial conditions $x_{1}=0.02$, $x_{2}=0.00$, $x_{3}=-0.10$ and $x_4=0.00$ in the second parameter setting. All the three trajectories exhibit some degree of chaotic dynamics, even though these chaos is very hard to appreciate in all of the cases.}
\label{fig5}
\end{figure}

\begin{figure}[ht!]
\centering
\includegraphics[width=1.0\textwidth]{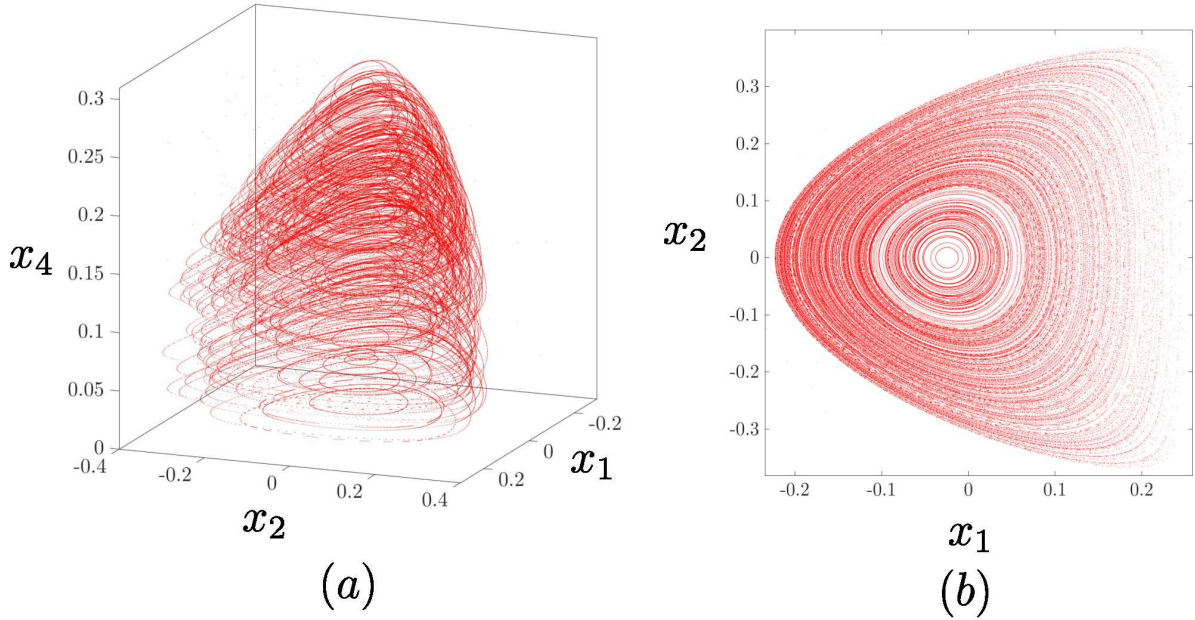}
 \caption{\textbf{Poincaré section and its projection}. (a) Using as 3D Poincar\'e section the hyperplane $x_3=0$, the Poincar\'e maps of randomly chosen orbits in the phase space are computed. Most trajectories seem to form circular rings of variable size, which bend in the 3D Poincar\'e section. (b) A projection of the Poincar\'e section onto the plane $x_1x_2$ is shown. Most of the orbits seem to organize around a common centre.}
\label{fig6}
\end{figure}

In Fig.~\ref{fig5} we show the projections of the phase space trajectories on the $x_1x_2x_3$ hyperplane, together with their associated Poincar\'e maps computed at the Poincar\'e section $x_3=0$.  These maps are obtained by computing the successive intersections of the trajectory with the hyperplane $x_3=0$. The resulting points $(x_1,x_2,0,x_4)$ are then projected on the $x_1x_2x_3$ hyperplane, for the three waves previously inspected. Naturally, the orbits in the Poincar\'e section unfold when represented in $x_4$, bending to form three-dimensional rings (see Fig.~\ref{fig6}(a)). We note that, even though the wave profiles always seem to exhibit some degree of sensitivity to initial conditions, their Poincar\'e maps are not very different from the maps commonly appearing in quasiperiodic motions. Therefore, it is evident that, if there is certainly chaotic dynamics, as some of the time series suggested, this dynamics is rather weak for the wave patterns here shown. This evidence is further confirmed when the 3D Poincar\'e section is fully computed, as depicted in Fig.~\ref{fig6}(a). Its projection onto the $x_1x_2$ plane exhibit as series of concentric rings Fig.~\ref{fig6}(b), while the chaotic sea seems to be missing. This structure certainly reminds of a KAM island. However, we note that the concentric rings belong to different values of $x_4$, and their structure is therefore not equivalent to a single KAM island. Given these facts, a systematic computation of the Lyapunov exponents is required to ascertain with more accuracy the chaotic dynamics of the system and its strength.

\section{Lyapunov Spectrum and chaotic sets}

We now explore the Lyapunov spectra for some of the patterns introduced in the previous sections. In the case under examination, which are nonlinear waves in plasmas, a positive Lyapunov exponent means that two waves profiles which differ at some point in space, separate eventually along their profile at some other points left behind or further ahead. This can be considered a signature of chaotic waves: if the profiles of the two waves are different at some point in space or time, no matter how small this difference is, the two profiles can not remain close everywhere, neither in space, nor in time. Moreover, the separation must increase exponentially fast.

\begin{figure}[ht!]
\centering
\includegraphics[width=1.0\textwidth]{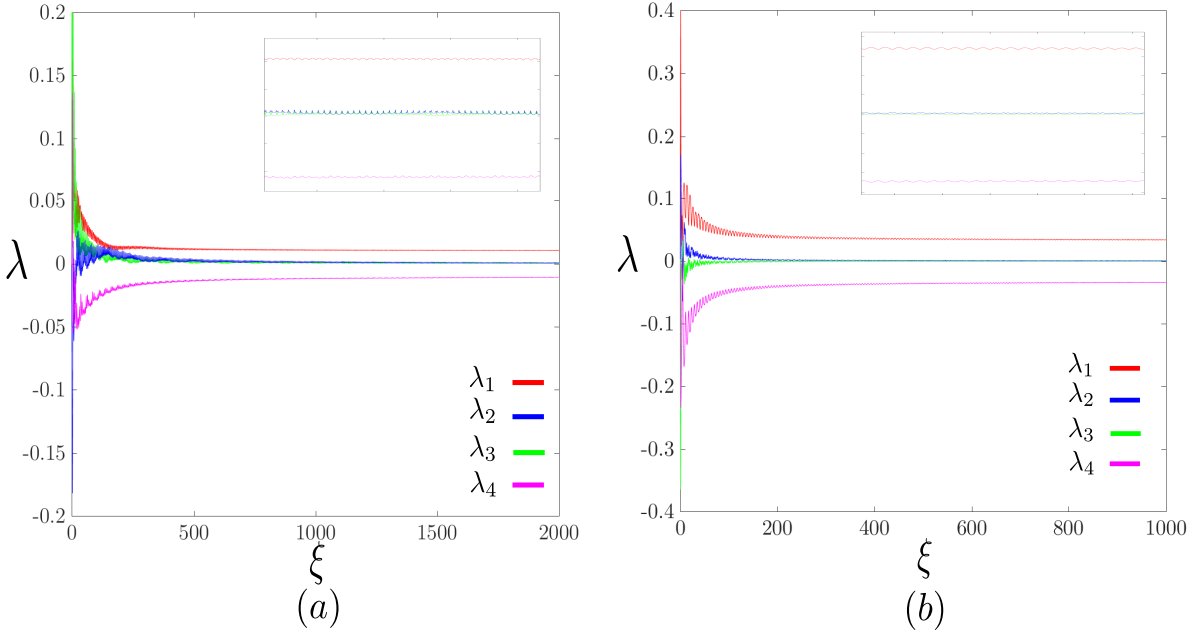}
 \caption{\textbf{Lyapunov spectra}. The whole Lyapunov spectrum has been computed for the two trajectories in the parameter setting $c=0.75$, $H=1.6$, $\rho=0.09$ and $N_d=0.98$. Two exponents are equal to zero and the remaining two have opposite sign. It can be seen that all the exponents add to zero, what confirms the average conservative nature of the system. The magnification appearing in the boxes show the fluctuations of the characteristic exponents around their asymptotic values. (a) The spectrum with initial conditions $x_1=-0.15$, $x_2=0.00$, $x_3=0.50$ and $x_4=0.10$ at $\xi=0$. (b) The spectrum with initial conditions $x_1=0.01$, $x_2=0.01$, $x_3=0.80$ and $x_4=0.01$ at $\xi=0$. This second case is more chaotic, since the maximum exponent has a value three times higher.}
\label{fig7}
\end{figure}

As can be seen in Fig.~\ref{fig7}, for the two waves appearing in Fig.~\ref{fig3} we have computed the whole Lyapunov spectrum. In the first case we got the Lyapunov exponents $\lambda_1=0.01$, $\lambda_2=0.00$, $\lambda_3=0.00$ and $\lambda_4=-0.01$. Thus we see that $\lambda_1$ and $\lambda_4$ are equal with opposite sign, while the two remaining are very close to zero both, confirming once more the average conservative nature of our dynamical system. Nevertheless, we note that very small fluctuations can be noticed by zooming in their asymptotic values, which are one order of magnitude smaller than their values. Thus the trajectories are weakly chaotic, but chaotic after all. As we get closer to the fixed point $x=(0,0,1,0)$, the chaotic behavior of the system enhances. When we consider the initial conditions of the wave appearing in Figs.~\ref{fig3}(c) and (d), the exponents are now $\lambda_2=0.03$, $\lambda_2=0.00$, $\lambda_3=0.00$ and $\lambda_4=-0.03$.

\begin{figure}[ht!]
\centering
\includegraphics[width=0.6\textwidth]{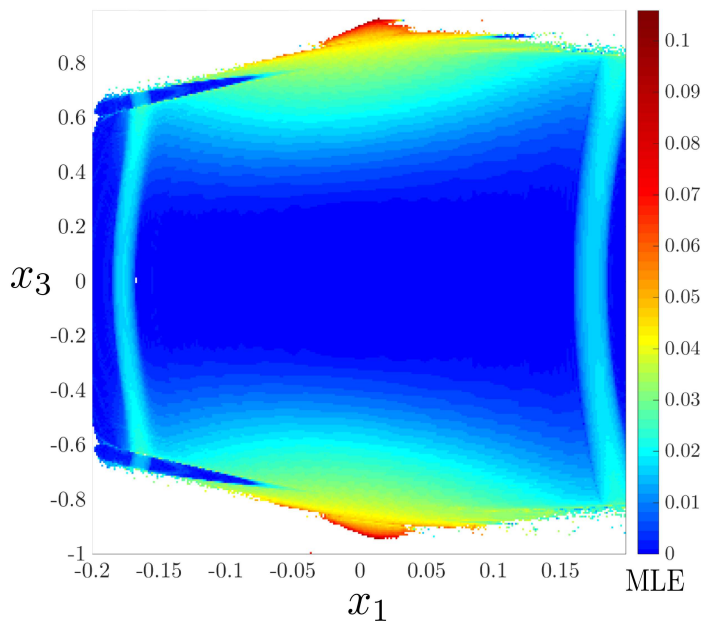}
 \caption{\textbf{Maximum Lyapunov exponent}. The MLE exponent has been computed for all the trajectories starting its journey at $x_2=x_4=0$, using the parameter setting $c=0.75$, $H=1.6$, $\rho=0.09$ and $N_d=0.98$. As can be seen, most of the regions exhibit very small values of the Lyapunov exponent (dark blue), while some hot regions with a non-negligible Lyapunov exponent are found for trajectories starting with very small ion speeds and high values of the electron density.}
\label{fig8}
\end{figure}

After these numerical findings, we now ask how strong can chaos be as we move over the phase space. For this purpose we explore the dependence of the maximum Lyapunov exponent (MLE) on the phase space coordinates. We select initial conditions from the plane $x_2=x_4=0.01$, using the original set of parameter values $c=0.75$, $H=1.6$, $\rho=0.09$ and $N_d=0.98$. Thus, we are considering perturbations of the particle's densities from equilibrium, with their derivatives equal to zero. As can be seen in Fig.~\ref{fig8}, while most of the region is dominated by a dark blue, where the Lyapunov exponents are too small to discard regular dynamics, some hot regions can be noticed as we increase the initial electron density towards one, while keeping small speeds of the ions $u_i$. In these regions the Lyapunov exponents acquire non-negligible values close to $0.1$, which means that it suffices to consider points far apart a value of ten units to observe deviations between the wave patterns. This numerical fact confirms the likely existence of chaotic dynamics in the present dynamical system.

To conclude our numerical explorations, we compute the chaotic parameter set for some fixed initial conditions. The chaotic parameter set of a dynamical system is defined as the value of the MLE computed for every pair of parameters in some parameter plane. We shall use the Mach number and the quantum parameter $(c,H)$ as the two parameters. When the system is ergodic, this set has a great intrinsic value, irrespective of initial or boundary conditions. However, in multistable systems or even Hamiltonian conservative systems, different initial conditions can have a different MLE. In our case, we will restrict to a fixed value at $\xi=0$, and explore two different points at this value. The first case corresponds to ion speed $x_1=0.17$ and a square root of electron density $x_3=0.20$, while their variations assume very small values of $x_2=0.01$ and $x_4=0.01$ as initial conditions.
\begin{figure}[ht!]
\centering
\includegraphics[width=1.0\textwidth]{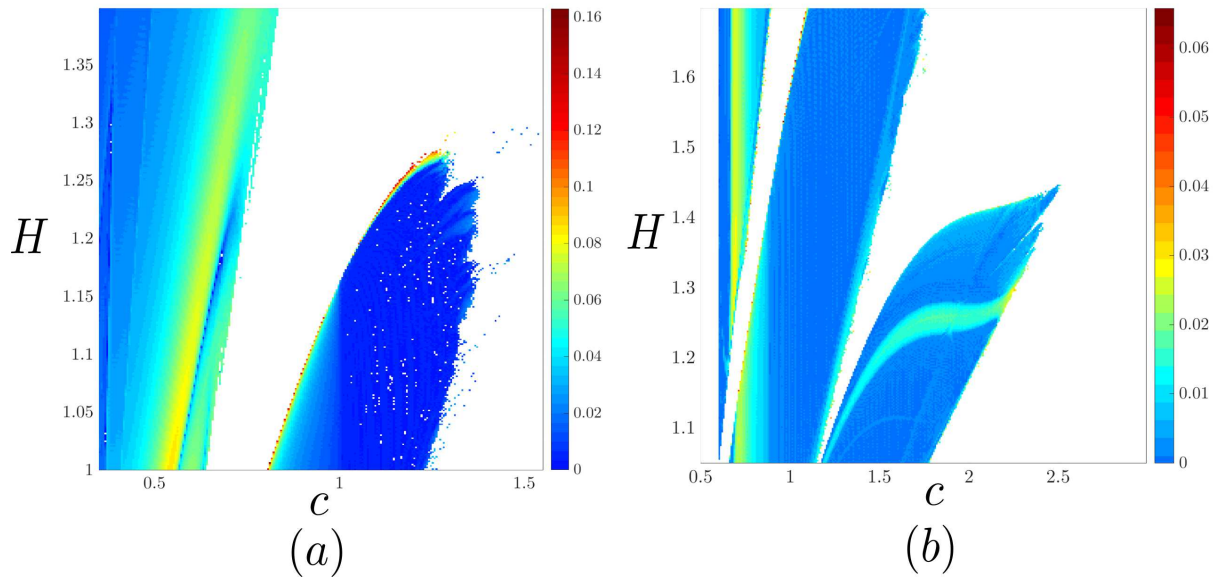}
 \caption{\textbf{Chaotic parameter sets}. The MLE is computed for a fixed initial condition at $\xi=0$ and letting the Mach number $c$ and the quantum parameter $H$ vary continuously in some domain. (a) A trajectory with initial condition $x_1=0.01$, $x_2=0.01$, $x_3=0.78$ and $x_4=0.01$. (b) A trajectory with initial condition $x_1=0.17$, $x_2=0.01$, $x_3=0.20$ and $x_4=0.01$. The colorbar represents the value of the MLE. As we can see, colored regions with bounded dynamics mostly exhibit small MLE, even though some hot regions are clearly detected at their boundaries.}
\label{fig9}
\end{figure}
As can be seen in Fig.~\ref{fig9}(a), we can see a wide colored region for small values of the Mach number, which harbors chaotic dynamics for some values of $c$. For all values of the quantum parameter $H$, the dynamics ceases to be bounded (uncolored) as we further increase the phase speed of the waves. Then, if the value of the quantum parameter is not too high, an elongated island where bounded dynamics reappears can be seen. The dusty regions have MLE very close to zero, but smaller than zero. We attribute these small deviations to the numerical scheme, which are probably associated to the long tails appearing in the distribution shown in Fig.~\ref{fig2}. Even though most of the chaotic regions in the parameter set have a MLE that is small, on the left shore of the blue island the dynamics of some regions where considerable values appear can be clearly appreciated. The second case corresponds to a high density of electrons $x_2=0.78$ at $\xi=0$ and small or negligible values of the remaining coordinates $x_1=0.01$, $x_3=0.01$ $x_4=0.01$. It displays not so different structure when compared to the previous case, but displaced to higher values of $c$.

\section{Discussion}

In the present work we have investigated a mathematical one-dimensional hydrodynamic model of a quantum dusty plasma. We have found as a novel result that the DIA waves exhibit the property of being conservative on average. Both analytical and numerical computations reveal the validity of this property, which means that the tendency of contractive effects in the plasma waves are counterbalanced by alternated transients of expansive dynamics. We wonder if during these spatiotemporal transients, intervals of hyperchaotic motion can be found, with two positive finite-time Lyapunov exponents. Importantly, since our demonstration can be extended to the limiting case in which the dust ions are absent ($N_d \rightarrow 0$), the property of ion acoustic waves of being conservative on average is also applicable to other type of waves in quantum plasmas.

Following previous works \cite{hass2003,ghosh2014}, we have explored the nature of several wave patterns and found out that some of them exhibit clear traces of chaotic dynamics. A dynamical phenomenon has been uncovered, by means of which, in some circumstances, the electrons tend to localize in some regions of the plasma, as a consequence of a comparatively small amplitude fluctuation in the ion concentration. This effect can lead to the depletion of electrons from widespread regions of the plasma, which might affect the physical properties of quantum well semiconductors. To delve deeper into the chaotic dynamics and acquire confidence in its occurrence, a systematic exploration of the Lyapunov spectra has been carried out for different wave profiles. Our results clearly indicate the existence of chaos in quantum plasmas, confirming previous research \cite{hass2003,ghosh2014}. Then, as far as the authors are concerned, the present work has attempted for the first time to investigate how chaotic dynamics of dust ions acoustic waves appears in the parameter space spanned by the Mach number and the quantum parameter. For this purpose, the chaotic parameter set has been computed, revealing regions where weak and not so weak chaos appears.

A heuristic argument in favor of the presence of chaos can be given as follows. The Eq.~\eqref{eq:23}, which relates the quantum force guiding the electrons through the plasma, together with the degeneracy pressure, lead to a differential equation that presents the typical shape of fundamental nonlinear oscillators with escapes, as for example the Helmholtz oscillator. The frequency of oscillation for the electron density is inversely proportional to the square of the quantum parameter and depends on the electrostatic potential. If we also include the other Eq.~\eqref{eq:26} describing the ion's speed, which also presents several nonlinear feedback terms and terms coupling to the electron dynamics, then, given the dimensionality of the dynamical system, we expect chaotic waves to be a fairly common event in general quantum plasmas.

To conclude, we highlight the fact that the ion quantum dynamics has been here disregarded, together with the magnetic and collision effects. We must also acknowledge that the effect of the electron exchange-correlation potential is important in the thermodynamic region at investigation. The influence of electron-exchange modifies the electron capture radius, charge capture process, capture probability, and capture cross section in degenerate quantum plasmas. Crouseilles \emph{et al.} were the first to introduce the exchange-correlation potential of degenerate electrons from density functional theory into the quantum hydrodynamic fluid equations in a phenomenological way, also discussing the limitations of this appromitation \cite{crouseilles2008}. This influence has been studied by sereval authors using a quantum hydrodinamical model \cite{mebrouk2014,mahmood2019,ahmad2020}, as well as through the kinetic theory approach \cite{zamanian2013,brodin2019}. Further investigation on how chaotic dynamics unfolds when all these features are considered is clearly deserved. In particular, as these phenomena and more species of charged particles are incorporated into a plasma, one wonders if typical features of hyperchaos and high-dimensional chaos are likely to arise.

\bibliography{mybibfile}

\end{document}